# Tailored Raman-resonant four-wave-mixing process

C. Ohae,[1,2] J. Zheng,[3] K. Ito,[3] M. Suzuki,[1,3] K. Minoshima,[1,2,3] and M. Katsuragawa[1,2,3,*]

[1]*Institute for Advanced Science, University of Electro-Communications, 1-5-1 Chofugaoka, Chofu, Tokyo 182-8585, Japan*
[2]*JST, ERATO, Minoshima Intelligent Optical Synthesizer Project, 1-5-1 Chofugaoka, Chofu, Tokyo 182-8585, Japan*
[3]*Department of Engineering Science, University of Electro-Communications, 1-5-1 Chofugaoka, Chofu, Tokyo 182-8585, Japan*

Nonlinear optical processes are strongly dominated by phase relationships among electromagnetic fields that are relevant to its optical process. In this paper, we theoretically and experimentally show, as a typical example, that in a Raman-resonant four-wave-mixing process, the first anti-Stokes and Stokes generations can be tailored to a variety of forms by manipulating the phase relationships among the relevant electromagnetic fields.

## I. Introduction

Nonlinear optical processes can be, in general, expressed using the conceptual illustration in Fig. 1. The nonlinear polarization, $P^{NL}$, is produced in the medium through mutual nonlinear interaction between the medium and electromagnetic fields that usually have different frequencies. Energy flows among such electromagnetic fields (energy flow from one mode in the electromagnetic fields to another) occur via this nonlinear polarization, $P^{NL}$, and these energy flows are strongly dominated by the phase relationships among the relevant electromagnetic fields, $\phi_a$, $\phi_b$, etc., including the phase of nonlinear polarization, $\phi_p$. This suggests that if it is possible to freely manipulate these phase relationships at various interaction lengths, we should be able to tailor this nonlinear optical process in a variety of ways [1]. This idea itself is general, and thereby it is possible in principle to apply this conceptual idea to a variety of nonlinear optical processes.

In the present study, we apply this conceptual idea to a Raman-resonant four-wave-mixing process and investigate it both theoretically and experimentally. Zheng and Katsuragawa Recently, it has been shown that incident electromagnetic field energy can be converted with near-unity quantum efficiency to an intended high-order anti-Stokes or Stokes mode by implementing these relative-phase manipulations in the Raman-resonant four-wave-mixing process in gaseous para-hydrogen [1]. In the same study, they also showed the potential to realize a single-frequency tunable laser that can cover an entire ultra-broad spectral region from 120 nm in the vacuum ultraviolet to the mid infrared. The main aim of this paper is to demonstrate a proof of concept in an actual medium and compare it with the above results that were obtained on the basis of theoretical and numerical investigations [2,3]. Since this was a first experimental trial, we focused our target on demonstrating a first order Raman-resonant four-wave-mixing process (the first step in Fig. 2a of Ref. 1),

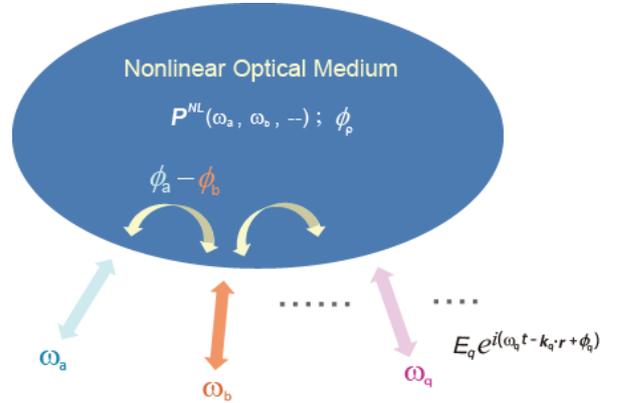

FIG. 1. Conceptual illustration illustrating the nonlinear optical process.

that is, the +1st anti-Stokes and Stokes generation processes.

Before proceeding with the main investigation, we briefly review former studies related to this work. As concerns tailoring nonlinear optical phenomena, various methods have been reported that design so called, phase matching, in a variety of forms. Quasi phase matching (QPM) [4-6] is one of the techniques, which, in recent years, has been extended to a new class of sophisticated QPM implementing functions in nonlinear optical processes. Various studies have been reported, including second harmonic generation with a broadband spectrum [7], engineered nonlinear optical-frequency conversions [8] by applying the concept of composite pulse [9,10], difference-frequency generation accompanied by shaped pulse structures [11], and nonlinear transverse-mode conversion from a Gaussian beam to a Hermite-Gaussian or Laguerre-Gaussian beam by extending the QPM into two dimensions by employing computer-generated binary holograms [12,13].

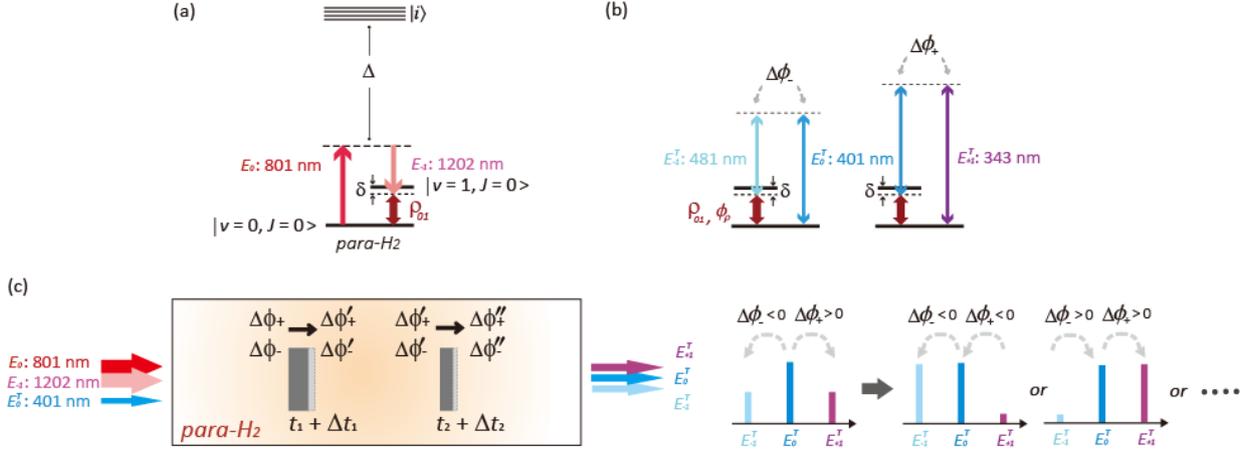

FIG. 2. Raman-resonant four-wave-mixing process implementing the manipulation of the phase relationships among the electromagnetic fields relevant to this nonlinear optical process. a, adiabatic driving (two-photon detuning: $\delta$) of vibrational Raman coherence, $\rho_{ab}$, in para-hydrogen molecule, where the driving laser radiations are $E_0$: 801 nm and $E_{-1}$: 1202 nm. b, 1st order Raman-resonant four-wave-mixing process: generations of 1st anti-Stokes, $E_{+1}^T$: 343 nm and 1st Stokes, $E_{-1}^T$: 481 nm, initiated from the incident third laser radiation, $E_0^T$: 401 nm. c, Conceptual configuration of the Raman-resonant four-wave-mixing process implementing a manipulation of relative phases ($\Delta\phi_+$, $\Delta\phi_-$ to $\Delta\phi'_+$, $\Delta\phi'_-$ and $\Delta\phi''_+$, $\Delta\phi''_-$) with transparent dispersive plates (thicknesses of the plates: $t_1+\Delta t_1$, $t_2+\Delta t_2$).

## II. Theory

Figure 2(a) shows an energy scheme for the Raman-resonant four-wave-mixing process investigated here. We introduce a pair of single frequency nanosecond pulsed laser radiations, $E_0$, $E_{-1}$, whose difference frequency is near resonant to the pure vibrational transition in gaseous para-hydrogen (ground state $|a\rangle$: $v = 0$, $J = 0$ to the vibrationally excited state $|b\rangle$: $v = 1$, $J = 0$), and drive the vibrational Raman coherence, $\rho_{ab}$, adiabatically from the ground state by controlling two-photon detuning $\delta$ [14-16]. States $|i\rangle$ are the far-off resonant intermediate states in this Raman transition ($\Delta$: far-off resonance frequency), which are dipole-transition-allowed electronically excited states. Here, we introduce another independent laser radiation, $E_0^T$. This third laser radiation, $E_0^T$, is phase-modulated by the produced vibrational Raman coherence, $\rho_{ab}$, which generates 1st anti-Stokes radiation: $E_{+1}^T$ and 1st Stokes radiation: $E_{-1}^T$ in the forward direction (Fig. 2(b)) [16,17].

To simplify the physical discussion, we reduce the amount of Raman coherence, $\rho_{ab}$, and effectively limit this optical process to the 1st order anti-Stokes and Stokes generations, although this nonlinear optical process is, in nature, open to high-order processes. We also set the amplitude, $E_0^T$, to be sufficiently lower than that of the driving laser radiations, $E_0$, and $E_{-1}$, to simplify the physical discussion by separating the optical process between the coherence driving process (Fig. 2(a)) and tailoring the 1st anti-Stokes and Stokes generations (Fig. 2(b)) [16,17]. Note that the efficiency of the Raman-resonant four-wave-mixing process is not, in principle, dependent on the incident energy of the third laser radiation [16,17].

This nonlinear optical process (coherent light-matter interaction) can be well described in the framework of the Maxwell-Bloch equation [14,18]. To see more explicitly how this light-matter interaction is dominated by the phase relationships among the electromagnetic fields relevant to this nonlinear optical process, we express the complex electric field, $E$, and Raman coherence, $\rho_{ab}$, as $E = \frac{\sqrt{\hbar\omega}}{\varepsilon_0}\sqrt{n}\exp(i\phi)$, $\rho_{ab} = |\rho_{ab}|\exp(i\phi_\rho)$, using photon-number-density, $n$, and phase, $\phi$ or $\phi_\rho$, respectively, and transform the formalism of the nonlinear coupled propagation equations to Eqs. 1 and 2 [1].

$$\frac{\partial\sqrt{n_{+1}}}{\partial\xi} = \frac{N\hbar|\rho_{ab}|}{\varepsilon_0 c}d_0\sqrt{\omega_0\omega_{+1}}\sin(\phi_{+1}-\phi_0+\phi_\rho)\sqrt{n_0} \quad (1)$$

$$\frac{\partial\sqrt{n_{-1}}}{\partial\xi} = -\frac{N\hbar|\rho_{ab}|}{\varepsilon_0 c}d_{-1}^*\sqrt{\omega_0\omega_{-1}}\sin(\phi_0-\phi_{-1}+\phi_\rho)\sqrt{n_0} \quad (2)$$

where $n_{+1}, n_{-1}, n_0$, $\omega_{+1}, \omega_{-1}, \omega_0$, and $\phi_{+1}, \phi_{-1}, \phi_0$ are photon-number-densities, angular optical frequencies, and phases of anti-Stokes (suffix, +1), Stokes (suffix, -1), and incident third laser radiation (suffix, 0), respectively. N is the molecular density, and $d_0$ and $d_{-1}$ are the effective dipole moments between states $|a\rangle$ and $|b\rangle$ for the 1st anti-Stokes and Stokes generations, respectively. To focus on the core physics, we neglect the terms corresponding to linear dispersions in Eqs. 1, 2.

As clearly shown in Eqs. 1, 2, the directions of energy flows (photon-number-density flows) among the electromagnetic-field modes in this nonlinear optical process are determined by the signs of the relative phases, $\Delta\phi_+ \equiv \phi_{+1} - \phi_0 + \phi_\rho$ and $\Delta\phi_- \equiv \phi_0 - \phi_{-1} + \phi_\rho$. When $\Delta\phi_+$ and $\Delta\phi_-$ take values between 0 - π and π - 2π, respectively, either $\sin(\Delta\phi_+)$ or $-\sin(\Delta\phi_-)$, respectively, becomes positive. The photon-number-density thereby flows from the incident laser radiation, $E_0^T$, to the 1$^{st}$ anti-Stokes, $E_{+1}^T$, and 1$^{st}$ Stokes radiations, $E_{-1}^T$: namely, the energies of the anti-Stokes and Stokes radiations increase with increasing interaction length, $\xi$. On the other hand, when $\Delta\phi_+$ and $\Delta\phi_-$ take values between π - 2π and 0 - π, respectively, either $\sin(\Delta\phi_+)$ or $-\sin(\Delta\phi_-)$ becomes negative, and thereby the photon-number-density flows in the opposite direction. More specifically, as the interaction length, $\xi$, is increased, the energies of the anti-Stokes, $E_{+1}^T$, and Stokes radiations, $E_{-1}^T$, are reduced, and conversely the energy of the incident laser radiation, $E_0^T$, is increased. These physical natures imply that it is possible to tailor this nonlinear optical process in a variety of ways if we can manipulate these relative phases, $\Delta\phi_+$ and $\Delta\phi_-$, to arbitrary values at a variety of interaction lengths in the nonlinear optical process.

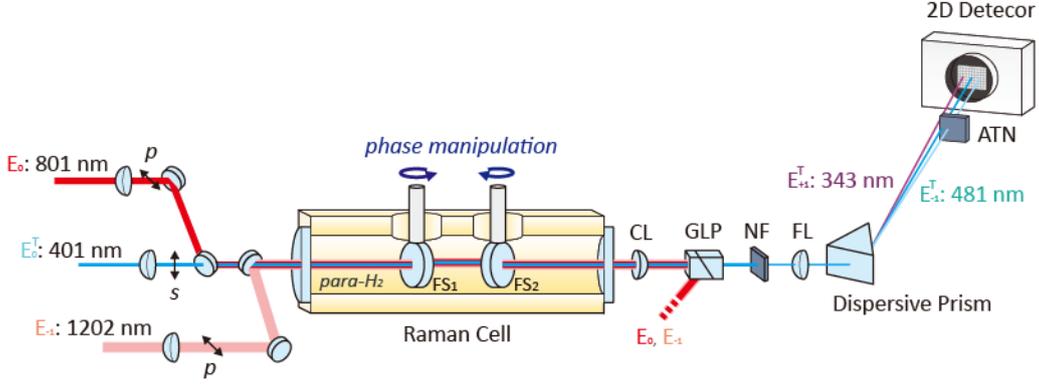

FIG. 3. Layout of the experimental system. CL: collimate lens, GLP: Glan laser polarizer, NF: notch filter, FL: focusing lens, ATN: attenuator. The entire system consists of lasers at three wavelengths (Raman-coherence driving lasers, $E_0$: 801 nm, $E_{-1}$: 1202 nm, and third laser, $E_0^T$: 401 nm), a copper-made Raman cell (interaction length: 340 mm) installed with a phase-manipulation device, and a detection system. The Raman cell is filled with gaseous para-hydrogen (density: $1.0 \times 10^{20}$ cm$^{-3}$). The two fused silica plates are also installed in the cell and their setting angles are controlled by the rotary stages placed outside the cell.

Here, the phase of the vibrational Raman coherence, $\phi_\rho$, is determined by the phases of the pair of nanosecond driving-laser radiations, $\tilde\phi_0$ and $\tilde\phi_{-1}$, simply expressed as $\phi_\rho = \tilde\phi_0 - \tilde\phi_{-1}$, when the produced Raman coherence is low and the high-order Raman-resonant four-wave mixing process can be neglected. Also, the relative phases, $\Delta\phi_+$ and $\Delta\phi_-$, between the incident laser radiation, $E_0^T$, and the anti-Stokes, $E_{+1}^T$, or Stokes radiation, $E_{-1}^T$, are intrinsically determined as being $+\frac{\pi}{2}$ and $-\frac{\pi}{2}$ via this $\phi_\rho$, respectively. In short, the energies of the electromagnetic fields (photon-number-densities) always flow in a unique direction from the incident laser radiation, $E_0^T$, to the anti-Stokes, $E_{+1}^T$, and Stokes, $E_{-1}^T$, radiations if we do not apply any phase manipulations and their initial phase relationships is also assumed to be maintained in a nonlinear optical medium.

When we examine the above-described conceptual idea (tailoring nonlinear optical process) in reality, the most difficult problem is how to achieve this manipulation of the relative phases at a variety of positions in a nonlinear optical medium. Surprisingly, we can find a very simple but very effective method by which to almost arbitrarily manipulate the relative phases. It is possible to tune phase relationships at will among the electromagnetic-field modes in question by inserting a transparent dispersive plate on an optical axis and controlling its thickness with a typical precision of several μm [19,20]; provided, however, that the frequency spacings among the relevant electromagnetic-field modes are required to be very wide (typically, greater than tens of THz). This optical method has been practically applied to a linear optical process, and has demonstrated the generation of a train of ultrafast pulses having a repetition rate exceeding 100 THz in the time domain [21-23]. Here, we apply this optical technology to tailoring nonlinear optical process: the Raman-resonant four-wave-mixing process [1,24,25]. We note that the point of this optical technology home in on an approximate solution. If we pursue an exact solution, however, this technology will no longer function practically [19,20].

### III. Experimental

The main part of the experimental system is illustrated in Fig. 3. We filled the copper cell (interaction length: 340 mm) with gaseous para-hydrogen (purity: > 99.9% [26], density: $1.0 \times 10^{20}$ cm$^{-3}$). This Raman cell was kept at room temperature. Under these experimental conditions, we introduced a pair of single-frequency nanosecond pulsed laser radiations [27] ($E_0$: 801.0817 nm, $E_{-1}$: 1201.6388 nm; linear polarization parallel to the optical table), whose difference frequency was near resonant with the vibrational Raman transition in para-hydrogen ($v = 1, J = 0 \leftarrow v = 0, J = 0$; resonant transition frequency: 124.7460 THz; two-photon detuning, $\delta$: 2.0 GHz) and drove the vibrational coherence, $\rho_{ab}$, adiabatically [14,15,16]. The excitation energies were 3.0 and 4.0 mJ for $E_0$ and $E_{-1}$, respectively, and the pulse duration was set to 10 ns. Each of the driving laser radiations was softly focused with a single lens, overlapped coaxially with each other, and introduced into the Raman cell. The beam diameters at the waist were set to 1.0 and 1.0 mm at $\frac{1}{e^2}$ for $E_0$ and $E_{-1}$, respectively.

Regarding the third laser radiation from which the Raman-resonant four-wave-mixing process was initiated, we employed the second harmonic ($E_0^T$: 400.5409 nm, linear polarization perpendicular to the optical table) of one of the driving laser radiations, $E_0$, by partially using its energy. The energy of $E_0^T$ (100 μJ) was set to be sufficiently lower than those of the driving laser radiations (3.0 or 4.0 mJ). As already described, this is because it enables us to investigate the main physical phenomena, namely, the Raman-resonant four-wave-mixing process initiated from the third laser radiation, $E_0^T$, independently of the adiabatic driving process of the vibrational Raman coherence, $\rho_{ab}$. This third laser radiation was then overlapped coaxially with the driving laser radiations and softly focused on the center of the Raman cell by a single lens. The beam diameter at the waist was set to 300 μm at $\frac{1}{e^2}$.

We also installed two transparent dispersive plates (5 mm-thick fused silica plates) on the optical path in the Raman cell, positioned 134 (FS$_1$) and 173 mm (FS$_2$) from the incident window, respectively, to tailor this Raman-resonant four-wave-mixing process. The fused silica plates were connected to rotary stages placed outside the cell, and their effective optical thicknesses were manipulated by controlling the insertion angles (resolution: 0.027 degrees). To prevent any influence of the incident energies of the driving laser radiations (parallel polarization) being reduced by the insertion of these fused silica plates, that is, having an influence on the vibrational-coherence driving process, we set the fused silica plates at the Brewster angle (56 degrees) and slightly manipulated their angles around it (maximum scanning angle: ± 2 degrees). The variation in the transmission loss of the driving laser radiations caused by scanning the insertion angles of the fused silica plates was minimized to less than 0.14% per plate.

The output from the Raman cell was spatially separated by a dispersive prism to the third laser radiation, $E_0^T$, and its 1st anti-Stokes, $E_{+1}^T$, and 1st Stokes radiations, $E_{-1}^T$. Each energy was then simultaneously measured on a two-dimensional detector. The two-wavelength driving laser radiations, $E_0$, $E_{-1}$, and their high-order stimulated Raman scatterings (horizontal polarization) initiated from the driving-laser-radiations themselves, which were also output coaxially with the third laser radiation, were eliminated by a Glan laser prism (GLP) placed after a collimate lens (CL). Because the output energy of the third laser radiation ($E_0^T$: 401 nm) was estimated to be much greater than those of 1st anti-Stokes and Stokes radiations, we also placed a notch filter with a central wavelength of 400 nm on the optical path of the third laser radiation, as depicted in Fig. 3, and adjusted all the signal levels at the three wavelengths, $E_0^T$, $E_{+1}^T$, and $E_{-1}^T$, to keep them within the linear-response dynamic range of the detector.

IV.  Results

As clarified in Eqs. 1, 2, the produced 1st anti-Stokes and 1st Stokes energies are strongly dominated by the phase relationships among the electromagnetic fields relevant to this nonlinear frequency mixing process, $\Delta\phi_+$ and $\Delta\phi_-$, respectively. In Fig. 4a, we plot the behaviors of $\Delta\phi_+$ and $\Delta\phi_-$ as a function of the effective optical-thicknesses of the inserted silica plates, from 8.75 to 8.95 mm, i.e., 54 - 58 degrees in terms of the insertion plate angle. For this overall manipulation of the plate thickness, $\Delta\phi_+$ (a-1) and $\Delta\phi_-$ (a-2) were varied by 6.9π and 4.4π, respectively. The directions of the photon number density flows in the 1st anti-Stokes and Stokes generations, determined by the signs of $\sin(\Delta\phi_+)$ and $\sin(\Delta\phi_-)$, are therefore expected to be reversed by 7 and 4 times, respectively. The reason that the periodicity is shorter in a variation of $\Delta\phi_+$ than that of $\Delta\phi_-$, is due to the optical nature of the refractive index dispersion of fused silica being greater at shorter wavelengths [28].

In advance of the experiment, we investigated as a numerical calculation, how the 1st anti-Stokes and Stokes generations can behave with this phase manipulation. We show the results in Fig. 4b. The regions colored in white represent that the 1st anti-Stokes or Stokes radiations were generated strongly (see the gray scale which is normalized by the maximum value). It was confirmed that the generated 1st anti-Stokes and Stokes energies varied nearly periodically as a function of the effective optical-thicknesses of the inserted fused silica plates, FS$_1$, and FS$_2$. This output-energy structures corresponded to the inversion periodicity (π) of the signs of $\Delta\phi_+$ and $\Delta\phi_-$, as shown in Fig. 4a, suggesting that behaviors of the 1st anti-Stokes and Stokes generations subjected to phase manipulation could be simply and well understood in terms of the directions of the photon-number-density flows.

By referencing the behavior of the 1st anti-Stokes and Stokes generations obtained in the numerical calculation (Fig. 4b-1,2), we conducted the corresponding experiment: tailoring of 1st anti-Stokes ($E_{+1}^T$: 343.3194 nm) and Stokes ($E_{-1}^T$: 480.6516 nm) generations, at 961 = 31 x 31 conditions by varying the insertion angles of the two fused silica plates from 54 - 58 degrees in 0.14 degree steps. As seen in Fig. 4c-1,2, the obtained experimental result accurately reproduced the behavior estimated in the numerical calculation.

To gain more details of their behaviors, we plotted in Fig. 4d-1,2 variations of the generated energies at 1st anti-Stokes (d-1) and Stokes (d-2) radiations against the optical thicknesses of the fused silica plate, FS$_1$, which are indicated in Figs. 4b-1,2, 4c-1,2 as white dotted lines. The details of these behaviors were affected by the following physical quantities: medium density, interaction length, refractive-index dispersion of para-hydrogen, positions and optical thicknesses of the inserted fused silica plates, and vibrational coherence produced. In the numerical calculations, each maximum value in the near-periodic variations in the generated anti-Stokes or Stokes energies closely approached each other when the produced vibrational-coherence, $\rho_{ab}$, was low. On the other hand, these maximum values varied against each other and the shapes of the near-maximal regions become steeper, as $\rho_{ab}$ became higher. The minimum value (offset) was also substantially affected by the refractive index dispersions of para-hydrogen. As it was difficult to apparently know the amount of vibrational coherence, $\rho_{ab}$, produced in the experiment, we treated it as an adjusted physical parameter in the numerical calculation, while we used the actual values estimated in the experiment for all the other physical quantities described above. The numerical calculations (gray solid lines) accurately reproduced the observed behaviors (blue dots) including the shapes of the periodic variations in the generated energies and the offset appeared in the minimal output energy regions. The produced vibrational Raman coherence in this experiment was estimated to be $\rho_{ab} = 0.0035$ by using this procedure, where the maximal conversion efficiencies at the 1st anti-Stokes and Stokes generations were 0.36% and 0.46%, respectively.

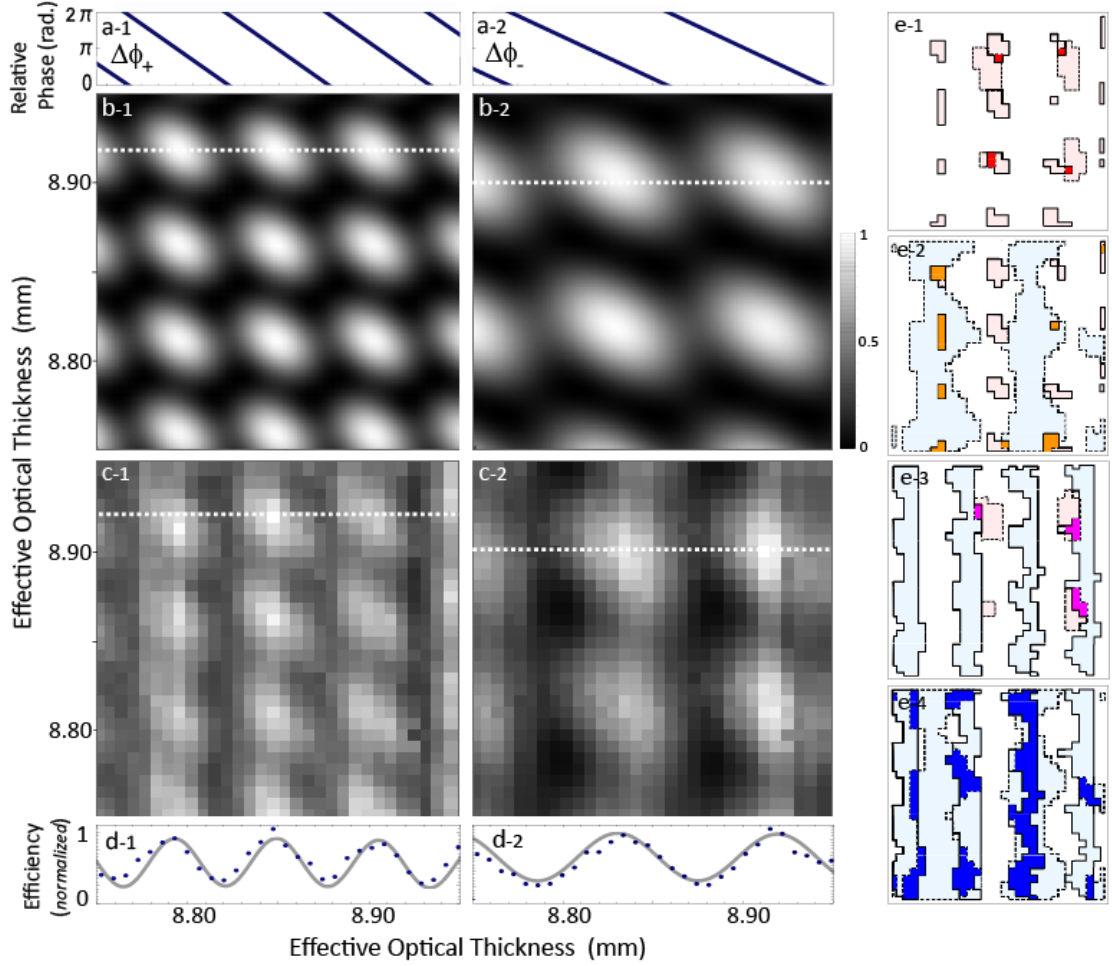

FIG. 4. Results of tailoring 1st anti-Stokes and Stokes generations in the Raman-resonant four-wave-mixing process by manipulating the phase relationships among the relevant electromagnetic fields. a, plots of the relative phases, $\Delta\phi_+$ (a-1) and $\Delta\phi_-$ (a-2) as a function of the effective optical thickness of a fused silica plate. b, (numerical calculation) counter plots of the anti-Stokes (b-1) and Stokes (b-2) energies generated against the effective optical thicknesses of the fused silica plates, $FS_1$ (horizontal axis), $FS_2$ (vertical axis), inserted in the Raman cell. c, counter plots of the anti-Stokes (c-1) and Stokes (c-2) energies observed for 961 (= 31 x 31) inserted angles of the two fused silica plates, $FS_1$, $FS_2$. The effective optical thicknesses corresponded to those in Fig. 4b-1,2. d, anti-Stokes (d-1) and Stokes (d-2) energies plot as a function of the effective optical thicknesses of the fused silica plate, $FS_1$: the generated anti-Stokes and Stokes energies are shown as white dotted lines in Figs. 4b-1, 4c-1 and Figs. 4b-2, 4c-2, respectively. The gray solid lines represent those obtained in the numerical calculation and the blue dots show those in the experiment. e, counter plots depicted by overlapping the anti-Stokes and Stokes energies obtained in Figs. 4b, c, where the plots are digitized into three levels (light red: region assigned above 80%, light blue: region assigned below 20%, against the full dynamic range of the generated energies). Four cases are shown: (e-1) both the anti-Stokes and Stokes, colored in red, were generated strongly; (e-2) only the anti-Stokes (orange) was generated strongly; (e-3) only the Stokes (pink) was generated strongly; and (e-4) both the anti-Stokes and Stokes (blue) were weak.

In fact, the 1st anti-Stokes and Stokes generations represented in Figs. 4c-1,2, respectively, occurred simultaneously. Figures 4e-1,2 comprise four types of counter plots depicted by overlapping them, whereas to accentuate the characteristic behaviors, we digitized them into three levels where the regions above 80% and below 20% against the full dynamic range of the observed output energies are colored in light red and light blue, respectively. The areas surrounded by solid and dotted lines respectively represent the results obtained for the 1st anti-Stokes and Stokes generations. It was demonstrated in the actual system that the Raman-resonant four-wave mixing process could be tailored in a variety of forms as exhibited in Figs. 4e-1-4: (e-1) both the 1st anti-Stokes and Stokes were generated strongly, colored in red; (e-2) only the 1st anti-Stokes was strong (orange); (e-3) only the 1st Stokes was strong (pink); and (e-4) they were both weak (blue) by precisely selecting combinations of the effective optical thickness (typical required precision: ~5 μm) of the two fused silica plates, FS$_1$, FS$_2$, that is, by simultaneously controlling the two relative phases $\Delta\phi_+$ and $\Delta\phi_-$.

## V. Conclusions

We have described the tailoring of nonlinear optical processes in a variety of forms by manipulating the phase relationships of the electromagnetic fields relevant to the nonlinear optical processes. We have demonstrated, as a typical example, that in the Raman-resonant four-wave-mixing process in gaseous para-hydrogen, the 1st anti-Stokes and 1st Stokes generations could be tailored to a variety of output energy combinations by applying phase manipulation method in which the effective optical thicknesses of fused silica plates inserted in the para-hydrogen cell were controlled. The observed results were also accurately reproduced by numerical calculation.

This experimental demonstration is regarded as a proof of principle for the study on tailoring high-order Raman-resonant four-wave-mixing processes which was investigated on the basis of numerical calculations in Ref. 1. The conceptual idea can be applied to a variety of nonlinear optical processes [25]. A single-frequency tunable laser covering the entire ultrabroad wavelength regions of vacuum ultraviolet to mid infrared as proposed in Ref. 1, as well as studies on the high resolution laser spectroscopy (in the vacuum ultraviolet and mid infrared region) based on this extreme laser technology should be considerable research interest.

*Email address: katsuragawa@uec.ac.jp

**Funding**. This work was conducted with the supports of a Grant-in-Aid for Scientific Research (A) No. 24244065, JST, ERATO Minoshima Intelligent Optical Synthesizer (IOS), JPMJER1304, and Gigaphoton Inc.

**Acknowledgment**. The authors thank Dr. K. Yoshii for his useful advice.